\documentclass[journal=jacsat,manuscript=article]{achemso}
\DeclareUnicodeCharacter{2212}{-}

\usepackage[version=3]{mhchem} 
\usepackage[utf8]{inputenc}
\usepackage{multirow}

\usepackage{float}

\usepackage{xcolor}
\usepackage{amsfonts,bm}

\newcommand{\kab}{k_{AB}}

\newcommand{\nmax}{N_\mathrm{max}}

\newcommand{\tcg}{T^\mathrm{clust}}
\newcommand{\wij}{w_{ij}}
\newcommand{\wzero}{w_0}
\newcommand{\xbf}{\bm{x}}

\usepackage{xcolor}

\author{Won Hee Ryu}
\affiliation[Oregon Health and Science University] {Biomedical Engineering, Oregon Health and Science University, Portland, OR, USA}

\author{John D. Russo}
\affiliation[Oregon Health and Science University] {Biomedical Engineering, Oregon Health and Science University, Portland, OR, USA}

\author{Mats S. Johnson}
\affiliation[Colorado State University] {Department of Mathematics, Colorado State University, Fort Collins, CO, USA}

\author{Jeremy T. Copperman}
\affiliation[Oregon Health and Science University] {Biomedical Engineering, Oregon Health and Science University, Portland, OR, USA}

\author{Jeffrey P. Thompson}
\affiliation[OpenEye, Cadence Molecular Sciences] {OpenEye, Cadence Molecular Sciences, Santa Fe, NM, USA}

\author{David N. LeBard}
\affiliation[OpenEye, Cadence Molecular Sciences] {OpenEye, Cadence Molecular Sciences, Santa Fe, NM, USA}

\author{Robert J. Webber}
\affiliation[California Institute of Technology] {Department of Mathematics, University of California San Diego, La Jolla, CA, USA}

\author{Gideon Simpson}
\affiliation[Drexel University] {Department of Mathematics, Drexel University, Philadelphia, PA, USA}

\author{David Aristoff}
\affiliation[Colorado State University] {Department of Mathematics, Colorado State University, Fort Collins, CO, USA}

\author{Daniel M. Zuckerman}
\affiliation[Oregon Health and Science University] {Biomedical Engineering, Oregon Health and Science University, Portland, OR, USA}

\email{zuckermd@ohsu.edu}

\title{Reducing Weighted Ensemble Variance With Optimal Trajectory Management}

\begin{document}


\begin{abstract}
Weighted ensemble (WE) is a path-sampling method that is conceptually simple, widely applicable, and statistically unbiased. In a WE simulation, an ensemble of trajectories is periodically pruned or replicated to enhance sampling of rare transitions and improve estimation of mean first-passage times (MFPTs). However, poor choices of the parameters governing pruning and replication can lead to high variance in MFPT estimates. Our previous work [J. Chem. Phys. 158, 014108 (2023)] presented an optimal WE parameterization strategy and applied it to low-dimensional example systems. The strategy harnesses estimated local MFPTs from different initial configurations to a single target state. In the present work, we apply the optimal parameterization strategy to more challenging high-dimensional molecular models, namely, synthetic molecular dynamics (MD) models of Trp-cage folding and unfolding, as well as atomistic MD models of NTL9 folding in high-friction and low-friction continuum solvents. In each system, we use WE to estimate the MFPT for folding or unfolding events. We show that the optimal parameterization reduces the variance of MFPT estimates in three of four systems, with dramatic improvement in the most challenging atomistic system. Overall, the parameterization strategy improves the accuracy and reliability of WE estimates for the kinetics of biophysical processes.

\end{abstract}


\section{1. Introduction} 

    With the fast-paced development of hardware and software, molecular dynamics (MD) simulation has become a common method to study the equilibrium and nonequilibrium properties of biophysical systems in atomistic detail.\cite{MD_1,MD_2,MD_3,MD_4,MD_5} However, MD faces a sampling problem. The integration time step in MD, usually a few femtoseconds, is many orders of magnitude smaller than the microsecond or longer timescales for biophysical processes of interest such as protein folding or allosteric transitions~\cite{sampling_problem_1,sampling_problem_2,trp_cage_Shaw}. The disparity in timescales makes it prohibitively expensive to obtain reliable estimates from direct MD simulation. To address this sampling problem, many enhanced sampling methods have been developed. Some widely used methods include umbrella sampling~\cite{eq_US_1,eq_US_2}, replica exchange~\cite{replica_exchange_1,replica_exchange_2,replica_exchange_3}, metadynamics~\cite{metaD_1,metaD_2}, coarse-grained models~\cite{CG_1,CG_2,CG_3}, transition interface sampling,~\cite{bolhuis2003novel,vanerp2017foundations} milestoning~\cite{faradjian2004computing}, forward flux sampling,\cite{forward_flux_sampling} and adaptive multilevel splitting~\cite{teo2016adaptive,brehier2016unbiasedness}.
    
    Weighted ensemble (WE) is an enhanced sampling method for equilibrium and nonequilibrium simulation that generates an ensemble of MD trajectories in parallel.\cite{huber_kim,WE_review} At a high level, WE allocates trajectories to regions of phase space that are valuable for computing observables of interest, for example, rarely visited barrier regions for conformational transitions. Each trajectory is assigned a weight, and WE manages the number of trajectories and their weights with regular pruning and replication, equivalent to the process of ``resampling'' in statistics\cite{Zuckerman_JCP_2010}.
    
\begin{figure}[t]
    \centering
    \includegraphics[width=\linewidth]{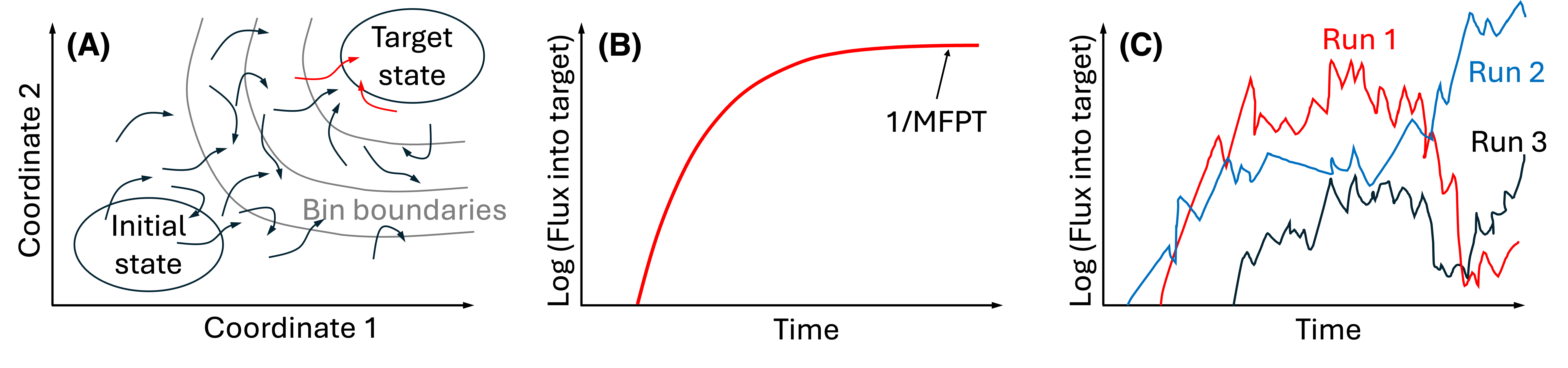}
    \caption{Rate constant estimation using WE (schematic). (A) The rate constant is estimated by the ``flux'' or trajectory weight entering the target state per unit time (red arrows). (B) Idealized flux curve over time for rate constant estimation from a WE run, which plateaus to a steady-state value equal to the reciprocal MFPT. (C) Realistic flux curves exhibit significant variance among different WE runs (run-to-run variance).} 
    \label{fig:we_rate}
\end{figure}

    Pruning and replication in WE are governed by {\em binning} and {\em allocation} hyperparameters. The binning parameters divide the phase space in selected coordinates (Fig.\ \ref{fig:we_rate}A), called ``progress coordinates'' or ``collective variables.'' The allocation parameters determine the target number of trajectories for each bin.  If a bin is occupied by fewer trajectories than the allocation, then trajectories are replicated until the allocation is reached. Conversely, if a bin's population exceeds the allocation, then trajectories are pruned. The trajectory weights are adjusted during replication and pruning to ensure unbiased estimation. Over time, the distribution of trajectory weights relaxes to the characteristic steady state of the system~\cite{Zuckerman_JCP_2010}. WE has been used to study a range of biophysical phenomena, such as protein conformational change,\cite{WE_protein_folding} protein-ligand binding,\cite{WE_protein_ligand_1,WE_protein_ligand_2} membrane permeation\cite{membrane_permeation}, protein-peptide binding, \cite{WE_protein_peptide_binding} protein-protein binding \cite{WE_protein_protein_binding}, and cryptic pocket detection\cite{cryptic_pocket}.

    WE is versatile and lightweight. It does not modify the underlying equilibrium distribution or the dynamics, unlike other enhanced sampling methods such as umbrella sampling or metadynamics.\cite{eq_US_1,eq_US_2,metaD_1,metaD_2} Additionally, WE performs replication and pruning of trajectories at fixed time intervals, in contrast to methods relying on interfaces such as transition path sampling,\cite{transition_path_sampling_1,transition_path_sampling_2} transition interface sampling,\cite{bolhuis2003novel,vanerp2017foundations} nonequilibrium umbrella sampling,\cite{noneq_US_1,noneq_US_2} milestoning,\cite{milestoning} forward flux sampling,\cite{forward_flux_sampling,sarupria2019contourFFS} and advanced versions of these techniques.\cite{peters2006aimless,cerou2011multiple,brotzakis2016oneway,brotzakis2019vietps,vanerp2003tis,moroni2004pptis,vanerp2005elaborating,vanerp2007pathswapping,swenson2014retis_multiinterfaces,hall2022retis_guide,warmflash2007neus,dickson2009neus_manyops,strahan2024badneus,vanden2009markovian_milestoning,bello2015exact_milestoning,ray2020wem,ray2022mwem,hussain2020ffs_review,kratzer2013ffs_interfaces,borrero2008ffs_optimizing,becker2012nsffs,elber2023firsthit,allen2006ffs_analysis,sarupria2019contourFFS}
    WE provides unbiased statistics for both equilibrium and nonequilibrium steady states, and it generates ensembles of trajectories leading to transitions.\cite{Zuckerman_JCP_2010,WE_kinetics_estimation_1}
    The fixed-time resampling scheme makes WE applicable to a range of dynamics engines for stochastic systems.\cite{Zuckerman_JCP_2010}

    Fig.\ \ref{fig:we_rate} shows how nonequilibrium WE is used to estimate the rate constant, defined as the reciprocal of the mean first-passage time (MFPT) for a transition process. The standard WE setup introduces a recycling boundary condition at the target state, leading to the reinitiation of trajectories at the initial state.\cite{huber_kim} This means that any trajectory reaching the target state (a single WE bin) is instantaneously recycled into the initial state (also a single bin) with its weight preserved. Once the WE trajectories reach the steady state, the average flux of trajectory weight into the target state determines the rate constant.\cite{hill_1} Although it is possible to estimate the rate constant during the transient relaxation period,\cite{hamsm_2020,WE_protein_folding} the present study will focus on steady-state WE simulation and analysis.

    While WE is motivated by simple ideas of replication and pruning, using WE to obtain accurate MFPTs can be a challenging -- if not prohibitively difficult -- task. 
    The high variance among independent WE replicates of the same system (run-to-run variance) makes results potentially unreliable,\cite{WE_protein_folding} as schematically shown in Fig.\ \ref{fig:we_rate}C. 
    Even though the mean of the estimated MFPTs is independent of WE hyperparameters, the run-to-run variance depends greatly on the hyperparameters. 
    For this reason, there has been a range of WE improvement efforts  based on milestoning,\cite{andricioaei2020WEM,andricioaei2021MWEM}  Gaussian accelerated MD,\cite{amaro2021gaMD-WE} machine-learned progress coordinates,\cite{leung_unsupervised_2025} binless strategies,\cite{dickson2019revo,dickson2022revo-ions,dickson2023revo-quality} alternative trajectory pruning strategies,\cite{ahn2024optimization} and post-analysis MFPT estimation methods.\cite{degrave2021red,hamsm_2020}
    Despite this body of work, the placement of WE bins has been largely ad hoc, without a systematic way to improve quality.
    Recent work has investigated adaptive bin placement along user-selected progress coordinates,\cite{MAB} yet this technique was designed to generate transitions, not minimize the variance of flux. 
    
\begin{figure}[t]
    \centering
    \includegraphics[width=\linewidth]{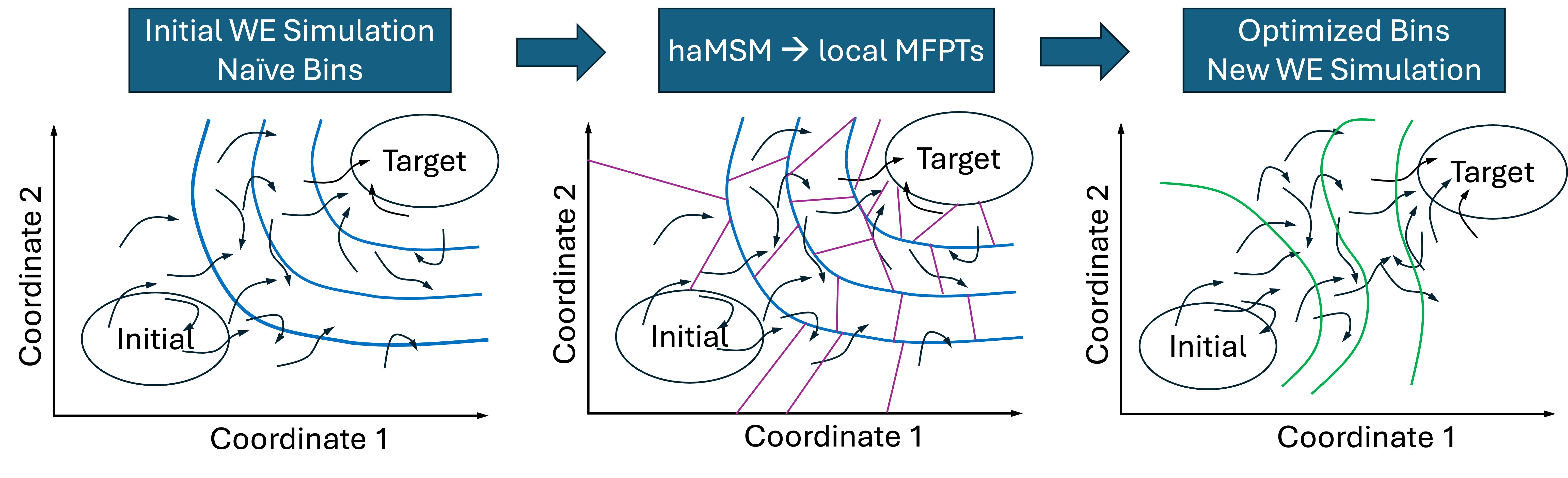}
    \caption{Optimizing WE bins. Training data for optimization is obtained from initial WE simulations using arbitrary bins (blue bin boundaries). After classification of phase space into states (purple and blue boundaries), a history-augmented Markov state model (haMSM) is constructed from the training data and used to estimate the local MFPT values for each state. The states are subsequently combined in an optimal way (green boundaries) to minimize variance in estimating the ``global'' MFPT from initial to target state.} 
    \label{fig:we_opt_scheme}
\end{figure}

    Here we build on our recent work that optimizes binning and allocation parameters to minimize the run-to-run MFPT variance.\cite{Aristoff_1,Aristoff_2,Aristoff_3,WE_recent_mathematical_developments} The approach, as outlined in Fig. \ref{fig:we_opt_scheme}, uses unoptimized WE trajectories (training data) to approximate a dynamical model for the system, called a ``history augmented Markov state model'' (haMSM).\cite{hamsm_2020} The haMSM is used to estimate two key quantities. The first quantity, denoted MFPT[$\xbf$], is the MFPT to the target state if the trajectories are all started from a specific starting configuration $\xbf$. \cite{schulten2003reaction,elber2020value} The second quantity is the variance of the local MFPT over a single interval of WE evolution. Our variance minimization strategy \cite{WE_recent_mathematical_developments} uses the local MFPT to group together similar trajectories so that groups of trajectories can be pruned without contributing excess error. The allocation contributes a high density of trajectories to high-variance regions, since additional sampling in these regions will improve the accuracy of the WE rate constant. The optimization scheme enables a systematic WE hyperparameter optimization in which the user only has to specify the total number of WE bins and the number of trajectories to maintain during WE.  

    Our previous study tested the WE optimization method on low-dimensional examples, leading to a $10,000\times$ improvement in the variance of estimated rate constants.\cite{WE_recent_mathematical_developments} However, it remained unclear how the method would perform in more realistic high-dimensional systems with complex free energy surfaces.
    
    In this work, we extend the variance minimization method to models with greater biophysical relevance, including a synthetic molecular dynamics\cite{synMD,trp_cage_Shaw} model of the Trp-cage miniprotein and an all-atom model of the N-terminal domain of ribosomal protein L9 (NTL9). When we apply variance minimization to Trp-cage folding and unfolding, it cuts down on outlier flux estimates, showing the efficacy of the approach. When we apply variance minimization to all-atom simulations of NTL9 folding in the low--friction setting, the efficacy is unclear. However, in the high-friction setting, applying variance minimization leads to significant improvements.
    
    The rest of this paper is organized as follows. In Section 2, we outline the underlying methods and theory for our approach. In Section 3, we describe the simulation details. In Section 4, we present results. Last, Section 5 provides discussion and Section 6 concludes.

\section{2. Methods and theory}

\subsection{Weighted ensemble simulation}

    WE has been described extensively in the literature \cite{huber_kim,Zuckerman_JCP_2010,WE_review}. In this section, we review the essential elements to set the stage for the optimization framework.

    \textit{Binning.} The WE method is based on a division of trajectories into non-overlapping groups called bins.
    WE can use time-varying bin definitions while remaining unbiased.\cite{Zuckerman_JCP_2010} However, for simplicity, the present study considers bins that remain static during a WE simulation.
    The bin boundaries are intervals of a user-specified progress coordinate, and a common choice is the root mean square deviation (RMSD) from a reference configuration. 

    \textit{Initialization.} The phase space is divided into $M$ bins with a target allocation of $n$ trajectories per bin, resulting in at most $\nmax = M \cdot n$ trajectories at any time. However, there can be $N < \nmax$ trajectories if there are any unoccupied bins.
    Each trajectory $k$ ($k=1,...,N$) at time $t$ is assigned a weight $w_t^k$, and the weights are normalized so that $\sum_{k=1}^{N} w_{t}^{k} = 1$. To initialize the WE simulation, all trajectories are started in a single bin, so there are just $n$ trajectories and the initial weights are $\wzero^k = 1/n$. 
    
    \textit{Dynamics.} Each trajectory is independently propagated according to the underlying dynamical model for a time interval $\tau$. The typical time interval is longer than a single time step but shorter than an MD simulation. 

    \textit{Resampling.}  Resampling is performed using the Huber-Kim procedure~\cite{huber_kim} implemented in the WESTPA software.~\cite{WESTPA_1,WESTPA_2} The trajectories arriving at an unoccupied bin are replicated or ``split'' into $n$ trajectories. If a bin contains more than $n$ trajectories, some are stopped based on a probabilistic resampling process\cite{Zuckerman_JCP_2010} with survival probabilities proportional to the weights: this process is called ``pruning'' or ``merging.'' The sum of the weights in each bin is maintained in the resampling process.
    
    \textit{Repetition.} The WE method alternates between dynamics and resampling steps until a target number of iterations is reached or until a user-chosen stopping criterion. 

\subsection{Rate constant and mean first-passage time via Hill relation}

    The effective first-order rate constant $\kab$ for a transition process from an initial state $A$ to a target state $B$ can be defined as the reactive flux, which is the fraction of the probability arriving at $B$ per unit time in the $A$-to-$B$ non-equilibrium steady-state ensemble. This definition is consistent with a chemical kinetics description $dP_B/dt = P_A \kab + \cdots$, which makes the simplifying assumption that $A$-to-$B$ transitions occur instantaneously.\cite{Zuckerman_textbook}

    It is convenient to represent our variance-optimization scheme using the MFPT instead of the flux.\cite{Aristoff_1,Aristoff_2,Aristoff_3,WE_recent_mathematical_developments} According to the Hill relation\cite{hill_1}, the MFPT equals the reciprocal of the steady-state flux $J$ from $A$ to $B$:
\begin{equation}
  \label{eqn:Hill_relation}
  \mathrm{MFPT}[\rho_{A}] = \frac{1}{J} = \frac{1}{\kab}.
\end{equation}
    The notation $\mathrm{MFPT}[\rho_{A}]$ indicates that we are evaluating the MFPT when the trajectories are released from an initial distribution $\rho_{A}$ within the state $A$. 

\subsection{Variance minimization using discrepancy and variance}

    The optimal WE strategy for minimizing the MFPT variance\cite{Aristoff_1,Aristoff_2,Aristoff_3,WE_recent_mathematical_developments} combines two procedures: (a) pruning \emph{similar} trajectories to limit the total number of trajectories; and (b) replicating \emph{important} trajectories that are most valuable for reducing variance.

    To make this strategy rigorous, we introduce the ``discrepancy'' and ``variance'' functions.
    The discrepancy is defined as
\begin{equation}
  \label{eqn:disc_mfpt}
  h(\boldsymbol{x}) = \frac{\mathrm{MFPT}[\pi] - \mathrm{MFPT}[\boldsymbol{x}]}{\mathrm{MFPT}[\rho_{A}]}.
\end{equation}
    The discrepancy is the difference between the MFPT starting from the steady-state distribution $\pi$ and the local MFPT starting from a point source $\xbf$, normalized by the ``global'' MFPT starting from $\rho_A$. Thus, the discrepancy kinetically orders the phase space, and it can be used as a WE progress coordinate. Regions with low discrepancy have high local MFPT values, making them kinetically far from the target state $B$. Regions with high discrepancy have low local MFPT values, making them close to the target. We note that the discrepancy conceptually differs from the better known ``committor'' function~\cite{committor_1,committor_2,committor_3,committor_4}, although the two may be correlated for systems with simple free energy landscapes. 
    
    The variance function is defined as
\begin{equation}
  \label{eqn:var_continuous}
  v(\xbf)^2 = \frac{1}{\tau} \, \mathrm{Var}\left[ \, h_0(\bm{X}_{\tau}) \, | \bm{X}_0 = \xbf \right],
  \quad \text{where } h_0(\bm{y}) = h(\boldsymbol{y}) + 1_B(\bm{y}).
\end{equation}
    Here, $h_0(\bm{y}) = h(\bm{y}) + 1_B(\bm{y})$ is a modified discrepancy function, where $1_B(\bm{y}) = 1$ if $\bm{y} \in B$ and $1_B(\bm{y}) = 0$ otherwise. 
    Eq.~\eqref{eqn:var_continuous} describes the variance of $h_0$ values when a trajectory is propagated from the initial phase point $\xbf$ for a time $\tau$. A region exhibits low variance when the MFPT values in the region are nearly uniform, for instance, within a deep free energy basin. Conversely, a region exhibits high variance if the trajectories in the region undergo large stochastic changes toward or away from the target state $B$, for instance, near a free-energy barrier. Intuitively, $v$ measures how much the local MFPT varies over one lag time, and the extra term $1_B$ is included to capture the flux into $B$.
      
    We optimize the bins and the allocation as follows. First, we generate bins using intervals of $h$ values: $h \in [h_{\rm low}, h_{\rm high})$, where we choose the endpoints of each interval so each bin contributes a constant integral $\int_{h(\bm{x}) \in [h_{\rm low}, h_{\rm high})} \pi(\bm{x}) v(\bm{x}) d\bm{x}$. 
    Each bin contributes equally to the variance of the local MFPT into $B$, and we allocate the same number of trajectories to each bin.
    In this way, we increase the density of trajectories in high-variance regions, improving the overall accuracy in estimating the MFPT.

    We can compare the optimal bin boundaries in WE to the optimal boundaries in transition interface sampling (TIS) and related methods\cite{transition_interface_sampling,bolhuis2003novel,vanerp2017foundations,cerou2011multiple}.
    These methods estimate the MFPT by directly observing paths to the target, so the ideal interfaces are isosurfaces of the committor function. 
    In contrast, WE averages the flux into the target over time, so the ideal interfaces are isosurfaces of the local MFPT.
    Also, rather than enforcing uniform progress as in TIS, optimized WE enforces uniform variance.
    
    In the next section, we will discuss the behavior and significance of the optimal allocation function $\pi v$, drawing on a one-dimensional (1D) double-well example.

\subsection{Discrepancy and variance in a 1D double well}

\begin{figure}[t]
    \centering
    \includegraphics[width=0.5\linewidth]{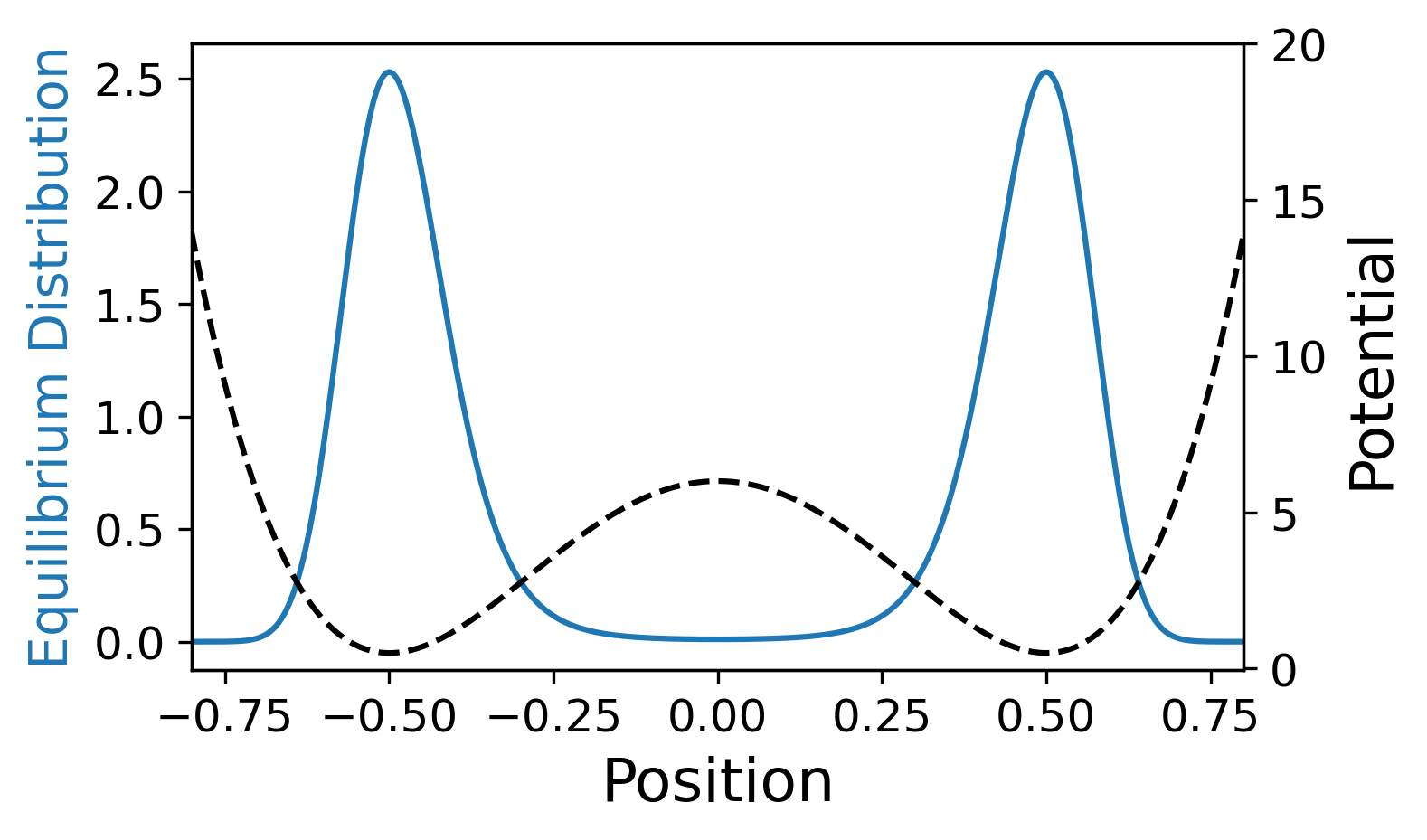}
    \caption{1D double-well potential function $V(x)$ and equilibrium density ${\rm e}^{-\beta V(x)} / \int {\rm e}^{-\beta V(y)}\mathrm{d}y$, plotted for $\beta = 1$. All quantities are dimensionless in this example.
    }
    \label{fig:1d_V_and_eq}
\end{figure}

    The WE optimization approach can be visualized through a 1D example based on the following unitless double-well potential function:
\begin{equation}
  \label{eqn:1d_potential}
  V(x) = 88 x^{4} - 44 x^{2} + 6.
\end{equation}    
    The system follows Brownian dynamics (also called overdamped Langevin dynamics):
    \begin{equation}
        dX_t = -V^{'}(x)dt + \sqrt{2/\beta} \,dW_t,
    \end{equation}
    where $W_t$ is a standard Brownian motion.
    Fig.~\ref{fig:1d_V_and_eq} illustrates the potential function and the associated equilibrium density ${\rm e}^{-\beta V(x)} / \int {\rm e}^{-\beta V(y)}\mathrm{d}y$ with $\beta = 1$.

\begin{figure}[t]
    \centering
    \includegraphics[width=0.95\linewidth]{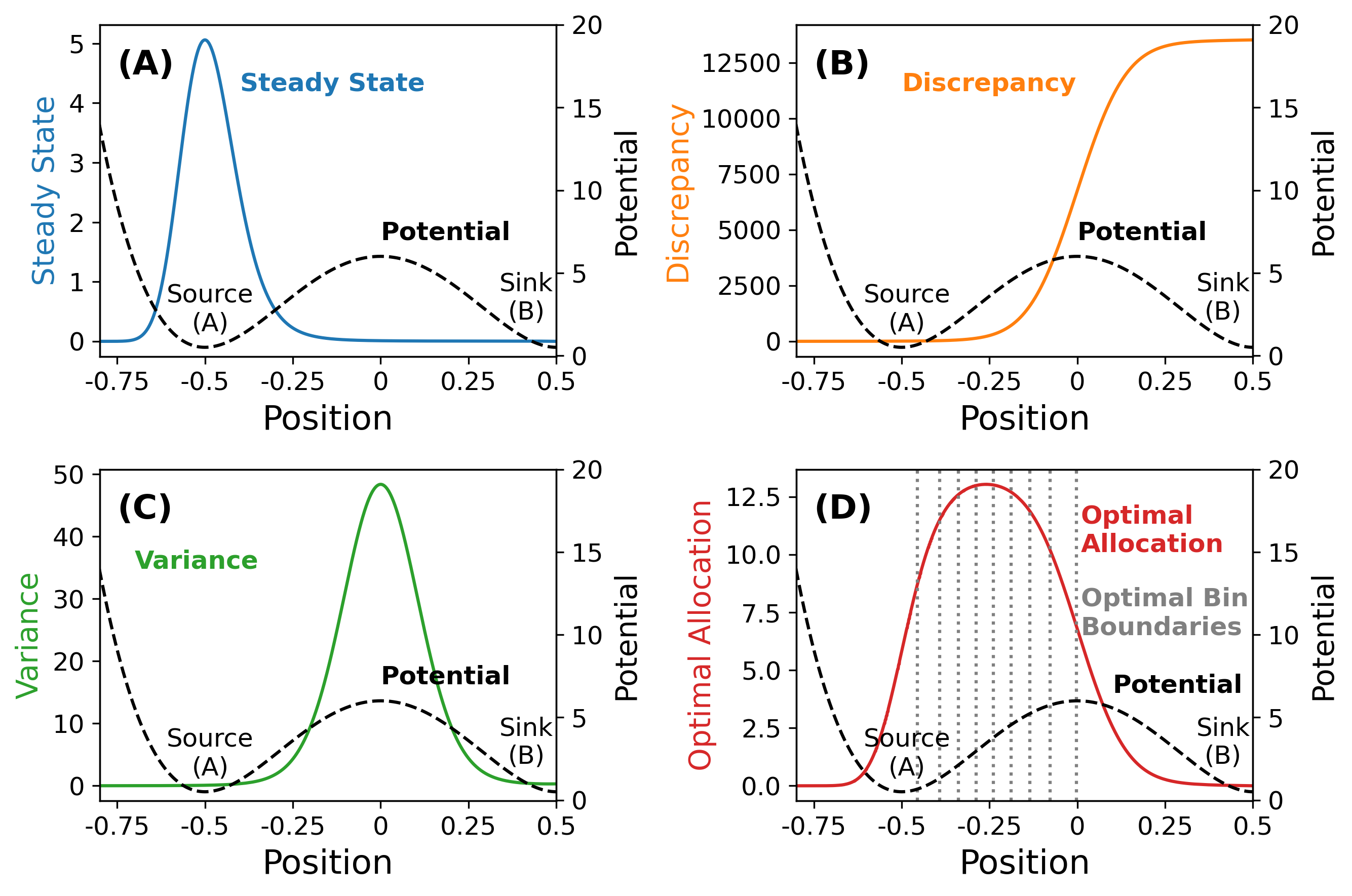}
    \caption{One-dimensional double well and various functions used in variance minimization, plotted with $\beta = 1$. The right edges show tick marks for the potential, while the left edges show tick marks for the other optimization functions. (A) Nonequilibrium steady-state distribution. The distribution $\pi$ (blue) is negligible near the sink, in contrast to the equilibrium steady-state distribution (Fig.\ \ref{fig:1d_V_and_eq}). (B) Discrepancy. The discrepancy $h$ (orange) illustrates that the local MFPT is larger near the source and smaller near the sink. (C) Variance. The variance $v^2$ (green) peaks near the energy barrier where trajectories might move to either side. (D) Optimal allocation. The optimal allocation $\pi v$ (red) is maximal on the ascending side of the barrier closer to the source. All quantities are dimensionless.
    } 
    \label{fig:1d_functions}
\end{figure}

    Next, we introduce ``recycling'' boundary conditions in which the trajectories that reach a target state $B$ are sent back to an initial state $A$. 
    Here the initial state $A$ is the point $x = -0.5$, and the target state $B$ is the region $\{x > 0.5\}$. By considering the continuous limit ($\tau \rightarrow 0$), we obtain closed-form equations for the non-equilibrium steady-state distribution, the discrepancy function, and the variance function:\cite{WE_recent_mathematical_developments}
\begin{align}
  \label{eqn:1d_steady_state}
  &\pi(x) \propto \int_{\max(x,A)}^{B} e^{\beta [V(y)-V(x)]}\mathrm{d}y, \\
  \label{eqn:1d_discrepancy}
  &h(x) \propto \mathrm{const} + \int_{-\infty}^{x}\int_{-\infty}^{z} e^{\beta [V(z)-V(y)]}\mathrm{d}y\,\mathrm{d}z, \\
  \label{eqn:1d_variance}
  &v(x) \propto \int_{-\infty}^{x} e^{\beta [V(x)-V(y)]}\mathrm{d}y.
\end{align}
    Figs.~\ref{fig:1d_functions}A-\ref{fig:1d_functions}C illustrate these functions with $\beta = 1$, leading to the following interpretations.
    \begin{itemize}
        \item Fig.~\ref{fig:1d_functions}A reveals that the non-equilibrium steady state $\pi$ has only one peak, whereas the equilibrium distribution shown in Fig.~\ref{fig:1d_V_and_eq} has two symmetric peaks. The recycling boundary conditions eliminate the second peak near the target state.
        \item Fig.~\ref{fig:1d_functions}B shows that the discrepancy function $h$ resembles a sigmoid function centered on the maximum of the potential energy barrier. The WE optimization approach uses this discrepancy function to kinetically sort the trajectories.
        \item Fig.~\ref{fig:1d_functions}C identifies that the variance function $v^2$ has a Gaussian shape centered at the top of the energy barrier. The WE optimization approach uses this variance function to identify regions for oversampling or undersampling.
    \end{itemize}

    The discrepancy function in Fig.~\ref{fig:1d_functions}B has similarities and differences from the well-known committor function. As a similarity, both functions quantify the progress of a trajectory going from the initial state to the target state. As a difference, the discrepancy is defined by the local MFPT (Eq.~\ref{eqn:disc_mfpt}) whereas the committor is defined by the probability of reaching the target state before reaching the initial state. Figure 5 of our previous work\cite{WE_recent_mathematical_developments} showed how these two functions differ in systems with multiple energy barriers: the discrepancy highlights the energy barrier with the greatest barrier height while the committor emphasizes the barrier that is farthest from the target state. Nevertheless, the committor and discrepancy are similar for the 1D double well (Eq.~\ref{eqn:1d_potential}) because only a single energy barrier is present.

    Last, Fig.~\ref{fig:1d_functions}D shows the optimal allocation function, which is the product of the steady-state density, $\pi(x)$, and the square root of the variance, $v(x)$. While the steady-state density quantifies the locations most frequently occupied by trajectories during a WE simulation, the variance quantifies the ``important'' regions where additional sampling can improve the accuracy of the flux estimate. In the 1D double well example, the optimal allocation leads to bins that are finely spaced on the uphill region leading to the energy barrier. Thus, the optimal allocation promotes a greater flow of trajectories toward the target state.

\subsection{Building MSMs from WE data: history-augmented MSMs}

    When exact formulas are unavailable, the WE optimization scheme requires estimates of the steady-state density, discrepancy function, and variance function. While it may be possible to use continuous approximations,\cite{covino} we choose to generate discrete approximations for these functions using Markov state models (MSMs).\cite{MSM_1,MSM_2,MSM_3} MSMs are already commonly used to estimate free energies and committor functions from MD trajectory data.\cite{MSM_free_energy_1,MSM_free_energy_2,MSM_committor_1,MSM_committor_2}

    Previous work has shown how the recycling boundary conditions in WE simulations make it possible to estimate nonequilibrium observables with less bias and shorter lag times than conventional MSMs.\cite{non_Markovian_1,non_Markovian_2,non_Markovian_3,non_Markovian_4} When a MSM is used in this way, it is called a ``history augmented MSM'' (haMSM). 
    The haMSM is constructed from WE data using a transition matrix $\bm{T}$ with entries \cite{hamsm_2020}
    \begin{equation}
    \label{eqn:tij_we}
        T_{ij} = \frac{\langle \wij \rangle}{\sum_j \langle \wij \rangle}.
    \end{equation}
    Here, $\wij$ is the total weight of trajectories transitioning from MSM cluster $i$ to $j$, and
    the angle bracket denotes an average over the WE iterations.
    We enforce the recycling boundary conditions by setting the row of transition probabilities starting from the target state equal to the row of transition probabilities starting from the initial state.

\subsection{Discrete formulation of discrepancy and variance}

    We use haMSMs to estimate the steady-state density $\pi$, the discrepancy function $h$, and the variance function $v^2$ according to the following discretization procedures~\cite{WE_recent_mathematical_developments}. 

    First, we calculate the steady-state vector $\bm{\pi} \in \mathbb{R}^n$ for a set of $n$ Markov states by solving the eigenvalue problem:  
    \begin{equation}
        \bm{\pi}^\top = \bm{\pi}^\top \bm{T},
        \quad \text{where } \sum_i \pi_i = 1.
    \end{equation}
    Here, $\bm{\pi}$ is the n-dimensional column vector that indicates the steady-state probabilities for all $n$ discrete states.

    Second, we calculate the discrepancy vector $\bm{h} \in \mathbb{R}^n$ by solving the linear system:  
\begin{equation}
  \label{eqn:disc_poisson}
  (\mathbf{I}-\bm{T})\bm{h} = 
\bm{T} \bm{1}_{B} - \pi_{B} \bm{1}.
\end{equation}
    Here, $\mathbf{I}$ is the $n \times n$ identity matrix;
    $\bm{1}$ is the $n$-dimensional vector of all ones; $\mathbf{1}_{B}$ has 1 entries indicating the target set and 0 entries otherwise; and $\pi_{B}$ indicates the steady-state probability corresponding to the target set $B$.
    
    Last, we set $\bm{h}_0 = \bm{h} + \mathbf{1}_{B}$ and calculate the variance vector $\bm{v}^2 \in \mathbb{R}^n$ by evaluating  
\begin{equation}
  \label{eqn:disc_var}
  \bm{v}^{2} = \frac{1}{\tau}\left[ \bm{T} \bm{h}_0^{2} - (\bm{T} \bm{h}_0)^{2} \right].
\end{equation}
    Here, $\bm{v}^2$ is the discrete variance of $\bm{h}_0$ with respect to $\bm{T}$, scaled by the inverse lag time $\tau$. The discrete formulation in Eq.~\ref{eqn:disc_var} matches the definition of variance for the underlying continuous dynamics in Eq.~\ref{eqn:var_continuous}.

\subsection{Optimized WE bin construction}

    Our optimized WE bin construction leverages two ideas: (a) the discrepancy function $h$ kinetically orders the phase space, and (b) the variance in the observed flux can be reduced by allocating more trajectories to regions of high local variance $v^2$. For the present study, we optimize WE bins using the following steps:
    
    \begin{enumerate}
        \item We compute $h$ and $v^2$ by building a haMSM model using preliminary simulation data obtained from a WE run with unoptimized bins. The haMSM model includes a substantially greater number of clusters than the number of WE bins.
        \item We order the phase space by increasing $h$ values.
        \item We calculate the integral of the allocation function $\pi v$ by using discrete sums based on the haMSM model. We set the bin boundaries so the sum $\sum_i \pi_i v_i$ is nearly the same for each bin, and we assign an equal number of trajectories to each bin.~\cite{WE_recent_mathematical_developments}
    \end{enumerate} \par

    By using the discrepancy function as a progress coordinate, we ensure that pruning some of the bins hardly impacts the future flux. By setting the number of trajectories in each bin proportional to $\pi v$, we are cutting down on the variance in the flux estimate. 
    
    We note that the WE bin construction has been optimized for rate constant and MFPT estimation with recycling boundary conditions, but it can be modified to target different observables in equilibrium settings.\cite{WE_recent_mathematical_developments} 
    
\subsection{Synthetic molecular dynamics (synMD)}
    
    To validate the variance reduction framework, we first apply it to a high-dimensional system with nontrivial behavior for which an exact reference solution is available. The synthetic molecular dynamics (synMD) system is a fine-grained MSM that associates each MSM state to an atomistic configuration.\cite{synMD} SynMD generates an MD trajectory with atomic coordinates that can be analyzed with standard post-MD analysis tools. Unlike conventional MD, however, exact results can be computed from the underlying transition matrix.

    In Sections 3 and 4, we will use the synMD model of Trp-cage folding and unfolding to build an ``exact coarse-graining'' (exact CG) transition matrix. Given a set of clusters indexed by $I$ and $J$, the exact CG transition matrix $\tcg$ has elements
\begin{equation}
  \label{eqn:exact_CG}
  \tcg_{IJ} = \frac{\sum_{i \in I}\sum_{j \in J} \pi_i \, T_{ij}}{\sum_{i \in I} \pi_i}.
\end{equation}
    Here, $T$ is the known microstate transition matrix, and $\pi_i$ is the steady-state probability of the $i$th microstate in the synMD model. The exact CG transition matrix in Eq.~\eqref{eqn:exact_CG} corresponds to a haMSM model in the limit of infinite sampling, and it helps us to assess whether there is sufficient data to accurately compute the cluster transition matrix.
    
\subsection{Hierarchy of phase-space discretizations}

    Before describing the simulation details, we summarize the levels of discretization of phase space in our study.
    
\begin{itemize}
    \item Coarsest level --- user-defined WE \emph{bins} (all systems). The ``bins'' describe the subsets of phase space used in WE simulation for splitting and merging. In this study we use both ``naive'' or unoptimized bins defined by RMSD and optimized bins defined by the variance reduction framework.
    \item Middle level --- data-driven MSM \emph{clusters} (all systems). The ``clusters'' describe the subsets of the phase space used for MSM analysis. 
    \item Finest level --- \emph{microstates} (synMD only). The ``microstates'' describe the subsets of phase space used in the discrete synMD model. When we apply MSM analysis to synMD trajectories, we group microstates into MSM clusters. However, we calculate exact reference properties using the underlying microstates.
\end{itemize}

\section{3. Simulation and system details}

    \subsection{Weighted ensemble software} 
        We performed all WE simulations with the WESTPA 2.0 software\cite{WESTPA_2} but used different MD engines to propagate the dynamics. We used a synMD module for the Trp-cage system, and we used AMBER for the NTL9 system.
        
    \subsection{Trp cage synMD model and WE parameters}

        The synthetic molecular dynamics (synMD) model was originally constructed\cite{synMD} by post-processing a 206 $\mu s$ equilibrium MD trajectory of the Trp-cage system generated by the Shaw research group.\cite{trp_cage_Shaw} SynMD is a fine-grained MSM with 10,500 microstates and a lag time of 1 ns, which we set to be the WE lag time. Each microstate is associated with an atomistic configuration, enabling coordinate-based computations.

        SynMD generates folding and unfolding events using a fixed microstate transition matrix. However, we applied different recycling boundary conditions for WE simulations of folding versus unfolding, where folded and unfolded states were defined by the RMSD of C$\alpha$ relative to the folded PDB structure (PDB ID 2JOF). We implemented the WE optimization by drawing data from 20 unoptimized WE runs, with each independent run lasting 2000 iterations.

        For Trp-cage unfolding, the folded state (source) was a single synMD state with RMSD of 0 Å (the PDB structure).
        The unfolded state (sink) was the set of synMD states with RMSD greater than 7 Å. The reference MFPT for unfolding calculated from the synMD transition matrix was 2.04 $\mu$s. 
        For the unoptimized runs, we placed bin boundaries at RMSD values of 0.0, 1.17, 2.33, 3.50, 4.66, 5.83, 7.0, and infinity Å. For the optimized runs, we updated the boundaries for the middle 6 bins but retained the first and last bins indicating the initial and target states.
        Each WE run used a large number of trajectories ($n = 20$) per bin.

        For Trp-cage folding, the unfolded state (source) was a single synMD state with RMSD of 12 Å.
        The folded state (sink) was the set of synMD states with RMSD smaller than 1.5 Å.  The reference MFPT for folding calculated from the synMD transition matrix was 9.03 $\mu$s. For the unoptimized runs, we placed bin boundaries at RMSD values of 0.0, 1.5, 2.0, 2.5, 3.0, 3.5, 4.0, 4.5, 5.0, 5.5, 6.0, 6.5, 7.0, 7.5, 8.0, 8.5, 9.0, 9.5, 10.0, and infinity Å. For the optimized runs, we updated the boundaries for the middle 18 bins but retained the initial and target bins, similar to the unfolding case. Each WE run used a fine discretization with 4 trajectories per bin. 
        
    \subsection{All-atom NTL9 simulation and WE parameters} 

        We simulated a continuous atomistic model of NTL9 using AMBER16 with the FF14SB force field\cite{AMBER_ff14SB} using NVIDIA GPUs (Titan-X and GTX-1080). Following our previous study,\cite{WE_protein_folding} we used the generalized Born implicit solvent model with two values of the friction coefficient, $\gamma = 5$ ps$^{-1}$ or $\gamma = 80$ ps$^{-1}$, which represent low-solvent viscosity (fast dynamics) and water-like viscosity (slow dynamics), respectively. We constrained the hydrogen bond length using the SHAKE algorithm and set the cutoff distance for nonbonded interactions to 99 Å. We defined folding in terms of the RMSD of C$\alpha$ relative to the reference folded structure (PDB ID 2HBB).
        
        When running the WE simulations, we used 54 bins with 4 trajectories allocated per bin. The unoptimized bin boundaries were uniform divisions of several intervals of the RMSD coordinate: 0.0 - 1.0 (1 bin), 1.0 - 4.4 (35 bins), 4.4 – 6.6 (12 bins), 6.6 – 10.2 (5 bins), and 10.2 – infinity Å (1 bin). As the only change from our previous study of NTL9,\cite{WE_protein_folding} we used a resampling interval of $\tau =$ 100 ps instead of $\tau =$ 10 ps.

    \subsection{Featurization and clustering for haMSM models and optimized bins}

        In this work, we used two types of featurization when building the haMSM models: C$\alpha$ distances and pair distances. C$\alpha$ distances are distances of alpha carbons from the current structure to the reference folded structure (same structure used to calculate RMSD), and pair distances are all possible pairwise distances of alpha carbons within a single structure. For example, given a protein with 10 alpha carbons, a featurization using C$\alpha$ distance would give a vector of length 10 whereas a featurization using pair distances would give a vector of length 45. 
        
        After constructing the two types of features, we performed dimensionality reduction using the variational approach for Markov processes (VAMP)\cite{VAMP_1,VAMP_2} with a variance cutoff of 0.95 and a kinetic map scaling. Then we constructed haMSM clusters in a hierarchical or ``stratified'' manner. Specifically, we assigned configurations to the appropriate WE bins and used the dimensionally reduced features to identify clusters within each bin. This hierarchical approach ensures that no haMSM cluster spans a wide range of RMSD values, which can happen when clustering is performed without stratification.
        The haMSM analysis was performed with an in-house code that can be found at \url{https://github.com/ZuckermanLab/WE_variance_minimization}.

    \subsection{Visualizing instantaneous WE flux data} 

        Regardless of the optimization scheme, the instantaneous WE flux data exhibits a high variance. In many of the iterations, there is only a small number of WE trajectories near the target state and there is no flux into the target. Additionally, when nonzero flux values do occur, they span multiple orders of magnitude because of repeated split and merge events. These characteristics make it challenging to visualize the instantaneous flux values. As a partial remedy, we visualized the flux for the Trp-cage synMD systems applied the following smoothing procedure based on moving averages.
        \begin{itemize}
            \item For each replicate in a group of WE runs, we identified the minimal window size to eliminate zeros in the moving average of the flux.
            \item Across all the window sizes for different replicates and different hyperparameters, we chose the largest value and used it uniformly for all comparisons.
        \end{itemize}
        We applied the moving average procedure for Trp-cage folding with a window size of 81 and Trp-cage unfolding with a window size of 185.
        Then we plotted the smoothed flux profiles with the selected window sizes in Figs.~\ref{fig:Trp_cage_unfolding}D and \ref{fig:Trp_cage_folding}D.
        
\subsection{Implementation and computing cost}

    The WE optimization procedure consists of three steps: (i) running initial, unoptimized WE simulations; (ii) building a haMSM model to compute optimized WE bins; and (iii) re-running WE with the optimized parameters. We implemented this workflow leading to the runtimes for NTL9 folding in Table \ref{table:computational_cost}.
        
\begin{table}[t]
\centering
\begin{tabular}{p{.19\textwidth}|p{.23\textwidth}|p{.21\textwidth}|p{.24\textwidth}}
\hline
Step in workflow & Result & Low-friction NTL9 & High-friction NTL9 \\
\hline
\multirow{4}{*}{Unoptimized WE} 
& 
\multirow{2}{*}{Amount of data}
& 5 replicates, \newline450 iterations each
& 20 replicates, \newline1000 iterations each \\
& Total molecular time 
& 32 $\mu$s 
& 277 $\mu$s \\ 
& Total runtime 
& 442 hours 
& 627 hours \\
\hline
\multirow{6}{*}{haMSM training} 
& 
\multirow{2}{*}{Amount of data}
& 5 replicates, \newline450 iterations each 
& 10 replicates, \newline1000 iterations each \\
& Dim.~ reduction
& $\sim$15 hours 
& $\sim$50 hours \\ 
& Stratified clustering 
& $\sim$7 hours 
& $\sim$60 hours \\ 
& haMSM construction
& $\sim$18 hours 
& $\sim$80 hours \\ 
& Total runtime 
& $\sim$40 hours 
& $\sim$190 hours \\
\hline
\multirow{5}{*}{Optimized WE} 
& 
\multirow{2}{*}{Amount of data}
& 5 replicates, \newline450 iterations each 
& 5 replicates, \newline1000 iterations each \\ 
& 
\multirow{2}{*}{Total molecular time}
& 16 $\mu$s (pairwise), \newline 37 $\mu$s (C$\alpha$) 
& 74 $\mu$s (3 clusters/bin), \newline 85 $\mu$s (4 clusters/bin) \\ 
& Total runtime
& \multicolumn{2}{c}{Similar to unoptimized WE per $\mu$s -- see text} \\
\hline
\end{tabular}%
\caption{Detailed runtime information for each step in the variance minimization workflow.}
\label{table:computational_cost}
\end{table}


The top row of the table shows that the runtime to generate 1 $\mu$s of unoptimized WE simulation data was $50\times$ faster in the high-friction than the low-friction setting.
Therefore we chose to generate more high-friction replicates (20 versus 5), each containing a greater number of iterations (1000 versus 450).
We used 10 of the high-friction replicates for haMSM training 
and used the rest only for the visual analysis in Figure~\ref{fig:ntl9_folding_high_friction}.



In our experiments, optimized WE ran $3 \times$ more quickly than unoptimized WE due to technical changes in the bin assignment function.
In unoptimized WE, the WESTPA software automatically assigned bins to each trajectory.
In optimized WE, a Python server saved the bin assignment function in memory, and a client script quickly returned the bin assignments.
An iteration of optimized WE with $n = 4$ trajectories took 13 seconds, whereas an iteration of unoptimized WE with the same number of trajectories took 41 seconds.
In the future, an updated WESTPA software could assign bins more quickly using a server-based approach, so unoptimized and optimized WE iterations would require a similar runtime.

\section{4. Results}

    This section first presents results for the synMD models of Trp-cage unfolding and folding, for which exact reference values are available. Then it presents results for the all-atom MD models of NTL9 folding in low-friction and high-friction solvents.

\subsection{Trp-cage unfolding (synMD)}

    \begin{figure}[H]
        \centering
        \includegraphics[width=0.95\linewidth]{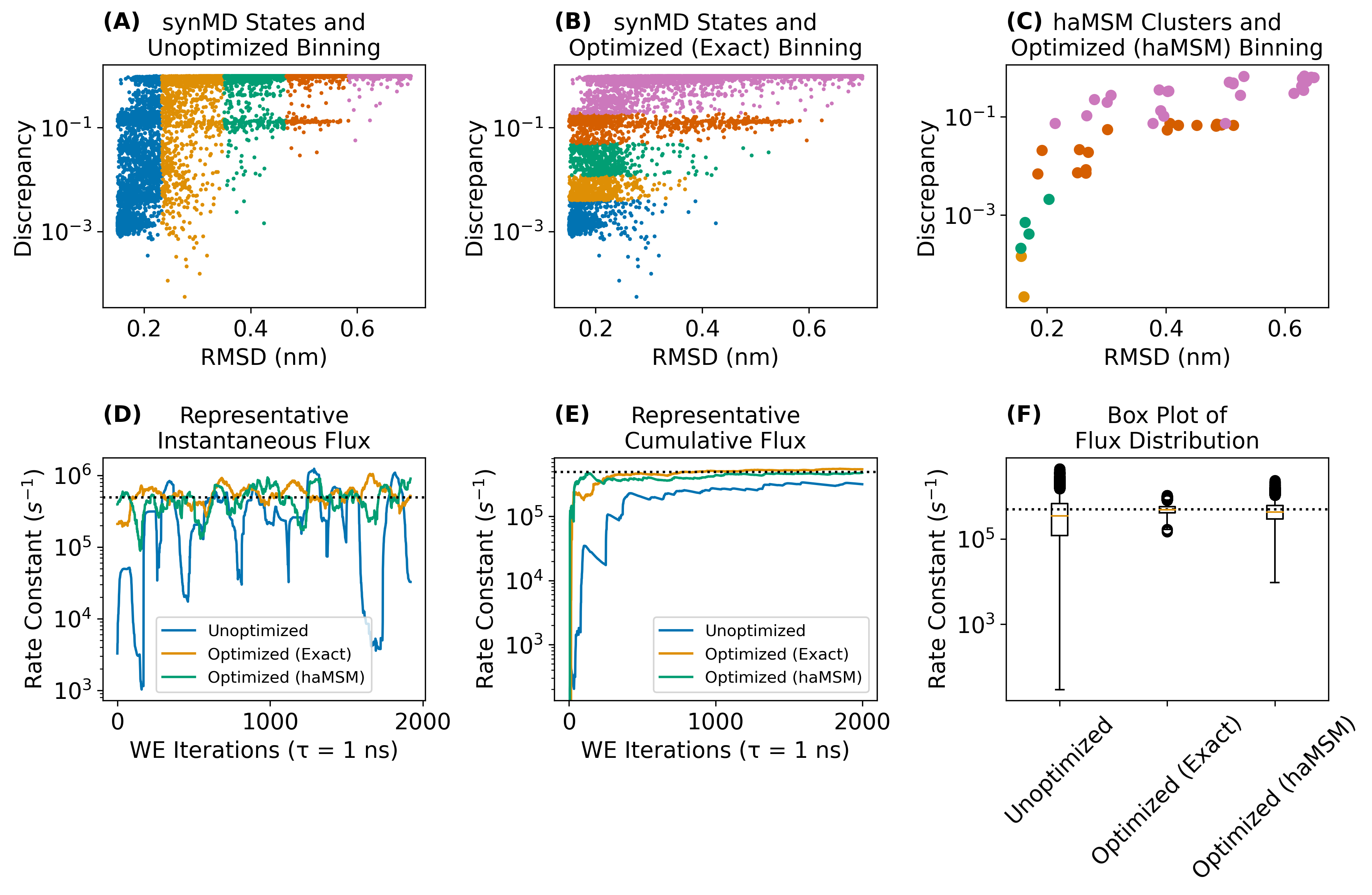}
        \caption{Variance reduction applied to WE simulations of synMD Trp-cage unfolding. The discrepancy functions in panels (A) to (C) have been rescaled to lie in the interval [0, 1] for visibility. (A) Unoptimized WE bins are indicated by colored bands in the RMSD coordinate, with synMD states plotted as dots. (B) Exactly optimized WE bins are indicated by colored bands in the discrepancy coordinate. (C) HaMSM optimized WE bins are indicated by color, with cluster centers plotted as dots. (D) Representative instantaneous flux profiles are compared to the exact reference value (dotted line). (E) Representative cumulative flux profiles are compared to the reference. (F) Box-and-whisker plots summarize the variation in flux values, based on 20 runs for each scheme. The horizontal orange lines indicate the median, while the top and bottom boundaries indicate Q3 and Q1 values, and the height of the box corresponds to the interquartile range (IQR). The top and bottom whiskers correspond to Q3 + 1.5 IQR and Q1 - 1.5 IQR, and the black circles correspond to outliers outside the whiskers.}
        \label{fig:Trp_cage_unfolding}
    \end{figure}

    We collected and compared flux data for the Trp-cage synMD system across three sets of WE bins: unoptimized, optimized (exact), and optimized (haMSM). Figs.~\ref{fig:Trp_cage_unfolding}A and~\ref{fig:Trp_cage_unfolding}B reveal the differences between optimized and unoptimized bins. In the unoptimized binning, the space is divided uniformly along the RMSD coordinate without accounting for the discrepancy information. Every RMSD bin contains a mix of microstates that are kinetically close and far from the target state. In contrast, the optimized binning schemes (both exact and haMSM) are based on the discrepancy function, so kinetically similar trajectories are grouped into bins.

    The exact and haMSM-based optimizations have two key differences: the resolution (microstate vs.\ cluster) and the strategy for calculating the transition matrix (exact vs.\ data-driven).
    The exact optimization uses the \emph{microstate} transition matrix to construct optimal bins, whereas the haMSM optimization uses unoptimized WE data to construct a haMSM model at the \emph{cluster} level in order to select near-optimal bins. The haMSM model is coarser than the microstate model and it indicates the level of resolution that could be realistically obtained from all-atom MD trajectories.

    The box plots in Fig.~\ref{fig:Trp_cage_unfolding}E summarize the flux estimates across 20 independent WE replicates.
    The comparison reveals that WE simulations with optimized bins exhibit smaller variations in flux --- the probability mass arriving at the target per unit time --- than the simulations with unoptimized bins.
    Also, the exact WE optimization produces even smaller variance than the haMSM-based procedure.

\subsection{Trp-cage folding (synMD)}

    \begin{figure}[H]
        \centering
        \includegraphics[width=0.95\linewidth]{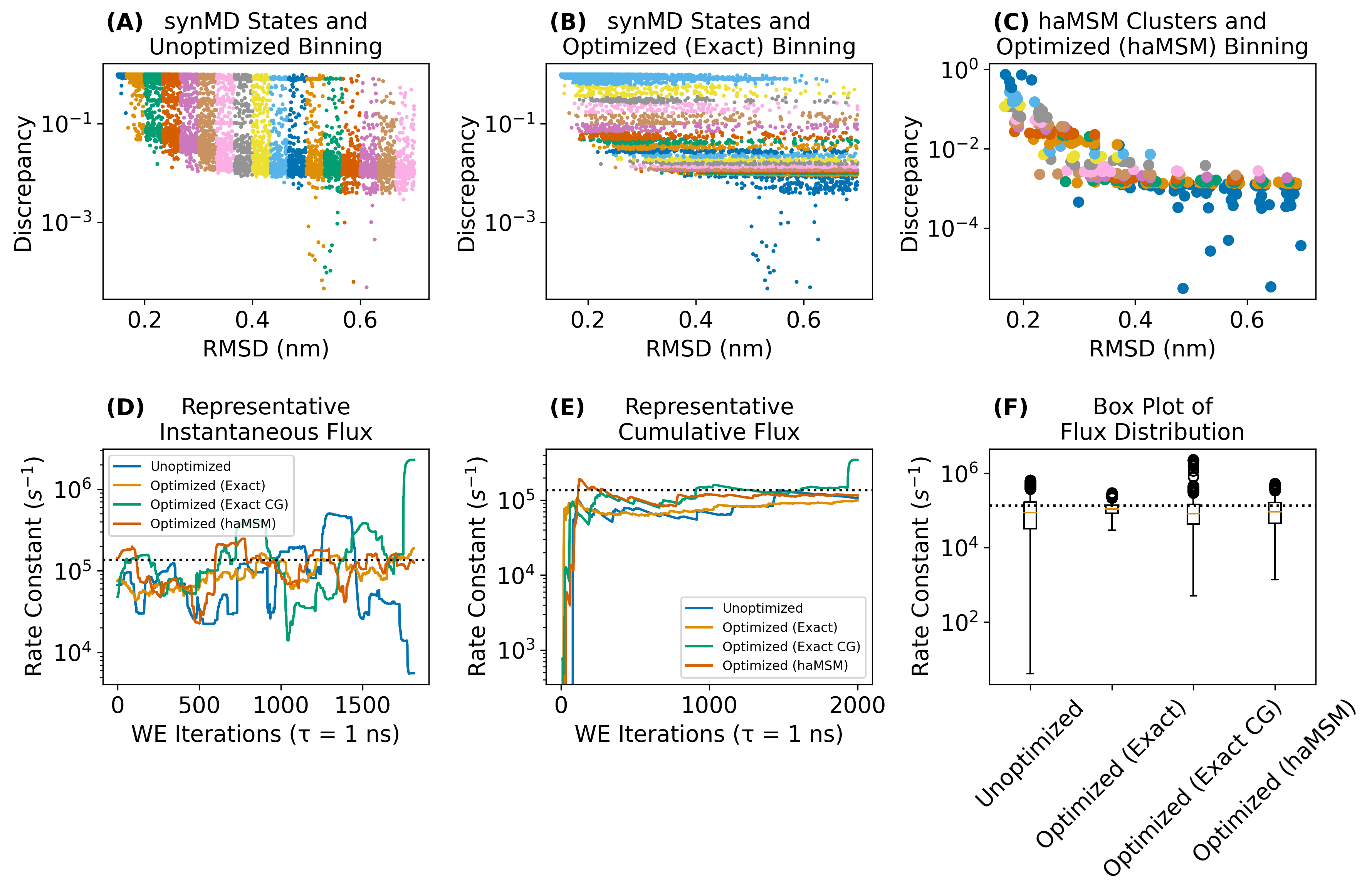}
        \caption{Variance reduction applied to WE simulations of synMD Trp-cage folding. The discrepancy functions in panels (A) to (C) have been rescaled to lie in the interval [0, 1] for visibility. (A) Unoptimized WE bins indicated as colored bands. (B) Exactly optimized WE bins indicated as colored bands. (C) haMSM optimized WE bins indicated by color, with cluster centers plotted as dots (D) Representative instantaneous flux profiles compared to the exact reference value (dotted line). (E) Representative cumulative flux profiles. (F) Statistical summary of variation in flux values based on 20 runs for each scheme.}
        \label{fig:Trp_cage_folding}
    \end{figure}

    Next we tested the WE variance minimization pipeline on the reverse process, Trp-cage folding. Because the results were more nuanced, we compared four WE binning approaches: unoptimized, optimized (exact), optimized (exact coarse-graining), and optimized (haMSM).
    
    Fig.~\ref{fig:Trp_cage_folding}A shows the unoptimized WE bins based on RMSD, which combine microstates with a wide range of discrepancy values.
    In contrast, Fig.~\ref{fig:Trp_cage_folding}B shows that exactly optimized WE bins are based on intervals of the discrepancy function, which ensure that the trajectories are kinetically similar.
    
    Fig.~\ref{fig:Trp_cage_folding}C shows the cluster centers assigned to each bin in the haMSM optimization procedure. The exact CG method uses the same cluster centers as the haMSM but it constructs different bins (not shown in the figure). In exact CG, the transition matrix is constructed in the limit of infinite data following Eq.~\ref{eqn:exact_CG}. Exact CG makes it possible to check whether inaccuracies in the data-driven transition matrix lead to poor WE bin construction.

    The box plots in Fig.~\ref{fig:Trp_cage_folding}F confirm that the WE optimization reduces the flux variance compared to the unoptimized WE simulations.
    Additionally, the variance of the exact CG and the haMSM bins are similar, which suggests that sufficient training data was used to construct the transition matrix. Nonetheless, the exact (microstate) bins produce lower variance than the exact CG or haMSM bins. Since the exact (microstate) bins have the highest possible resolution, 
    this comparison suggests that higher resolution corresponds to better performance in the WE optimization.

\subsection{Low-friction NTL9 folding}

\begin{figure}[t]
        \centering
        \includegraphics[height=0.4\linewidth]{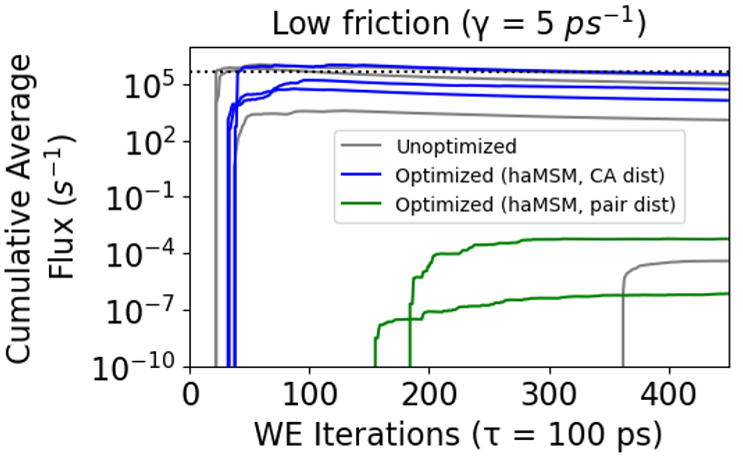}
        \caption{Cumulative average flux for WE simulations of low-friction NTL9 folding.  Gray lines show unoptimized WE runs,
        blue lines show optimized WE runs using features based on C$\alpha$ distances to the target, and green lines show optimized WE runs using features based on pairwise C$\alpha$ distances. The dotted black line shows the reference flux value from the previous literature~\cite{WE_protein_folding}.}
        \label{fig:ntl9_folding_low_friction}
    \end{figure}

    To further test the efficacy of the optimization pipeline, we applied our method to simulate the folding of NTL9, which we modeled using atomistic MD with either a low-friction or high-friction solvent. In this section, we focus on the low-friction results, and we evaluate the WE optimization using features based on either the C$\alpha$ distances to the target folded structure or the pairwise distances between alpha carbons. Fig.~\ref{fig:ntl9_folding_low_friction} summarizes the results of the low-friction NTL9 simulations. 

    Low-friction NTL9 was the only system examined in this study for which the WE optimization was not clearly effective. 4 of 5 unoptimized WE runs exhibited folding events. Meanwhile, 3 of 5 optimized WE runs led to folding events using features constructed from C$\alpha$ distances to the target.
    Just 2 of 5 optimized WE simulations exhibited folding events using features constructed from pairwise C$\alpha$ distances. Although 5 is a small sample size, the results suggest that the ``optimized'' WE runs can perform worse than the unoptimized runs.
    One possible explanation is that the the 100 ps lag time used in this study was not long enough to model the NTL9 dynamics in the low-friction solvent.
    
\subsection{High-friction NTL9 folding}

    Last, we tested the optimization scheme on a high-friction model of NTL9, which is characterized by a Langevin friction coefficient of 80 $ps^{-1}$ instead of 5 $ps^{-1}$. Once again, the model is based on true atomistic MD. The increased friction leads to slower relaxation times and higher run-to-run WE variance.\cite{WE_protein_folding}

    \begin{figure}[t]
        \centering
        \includegraphics[height=0.4\linewidth]{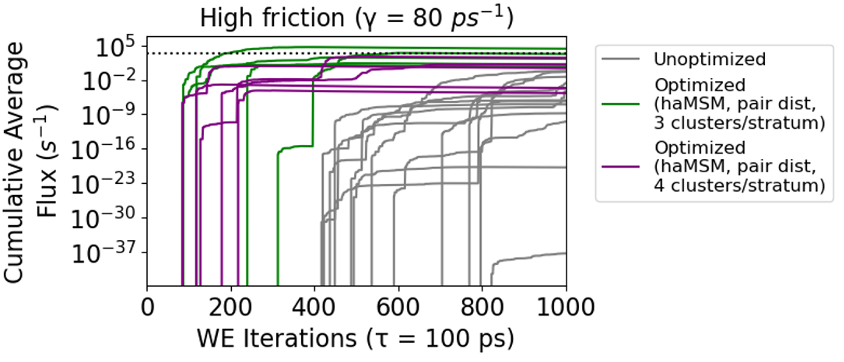}
        \caption{Cumulative average flux for WE simulations of high-friction NTL9 folding. Gray lines show unoptimized WE runs, green lines show optimized WE runs using 3 clusters per bin, and purple lines show optimized WE runs using 4 clusters per bin.
        The dotted black line shows the reference flux value from the previous literature~\cite{WE_protein_folding}.}
        \label{fig:ntl9_folding_high_friction}
    \end{figure}
    
    Fig.\ \ref{fig:ntl9_folding_high_friction} shows the benefits of the WE optimization for the high-friction system. Among 20 unoptimized runs, only 14 exhibited folding events. Meanwhile, 10 of 10 optimized runs yielded folding events and the variability in flux estimates decreased dramatically. 
    
    Last, we compared optimization strategies based on haMSMs with slightly different resolutions of 3 or 4 clusters per unoptimized WE bin.
    While the slightly lower resolution model (3 clusters/bin) exhibits better flux behavior (Fig.\ \ref{fig:ntl9_folding_high_friction}), this comparison is based on only 5 runs of each type and cannot be considered conclusive.

\section{5. Discussion}

    Weighted ensemble is both blessed and cursed by its flexibility. The method provides unbiased estimates of observables regardless of the binning scheme\cite{Zuckerman_JCP_2010}, but many binning schemes may not be sufficiently efficient for practical simulations. Due to the lack of a systematic framework, most biomolecular WE simulations have used hand-crafted binning schemes,\cite{WE_review} with user choices that are based on intuition and likely far from optimal. For example, in the protein folding systems studied here, RMSD is weakly correlated with the kinetic behavior embodied in the local MFPT (Figs.\ \ref{fig:Trp_cage_unfolding} and \ref{fig:Trp_cage_folding}).

    The present study shows that a previously developed optimization framework for WE binning, which had only been applied in low-dimensional systems\cite{WE_recent_mathematical_developments}, is indeed capable of improving performance for complex molecular models. In the optimization method, simulation data from unoptimized WE runs is used as training data to approximate the local MFPTs and local variances, which in turn define the optimal binning. In other words, starting from low-quality data, we can systematically improve the WE bins.
    
    In the slowest-relaxing atomistic system --- high-friction NTL9 folding --- the WE optimization method substantially reduced the run-to-run variance even though the training data exhibited a high variance. Six out of twenty unoptimized WE runs did not yield any folding events, while the remaining runs led to flux estimates spanning 35 orders of magnitude. In contrast, all ten optimized WE runs produced folding events, and the flux estimates spanned 12 orders of magnitude. The challenging nature of rate-constant estimation suggests the main goal should be obtaining the correct order of magnitude, and the WE optimization is clearly helpful.

    We initially studied synthetic MD systems which have exact reference solutions. One lesson learned from synthetic systems is that the haMSM model needs many clusters in suitable coordinates to effectively estimate the functions used for optimization. In some of our initial attempts at optimization, there were not enough clusters and the resulting WE bins were coarsely constructed. We leveraged this insight by building haMSMs for atomistic systems using the finest clustering which the available data would support.

    In this study, we employed hierarchical clustering by first stratifying on a known slow coordinate (RMSD). Otherwise, we have seen that the clusters --- which are supposed to represent kinetically similar configurations --- may span large ranges of the slow coordinate. 
    We explored additional clustering details in our atomistic NTL9 experiments. Yet the observed differences between pairwise and single-atom coordinates (Fig.\ \ref{fig:ntl9_folding_low_friction}) or the  differences based on the number of clusters per unoptimized WE bin (Fig.\ \ref{fig:ntl9_folding_high_friction}) are difficult to interpret due to the small sample size.

    Like any model, the haMSM model is only as good as the training data set. While one can perform sanity checks like lag time analysis \cite{mey2024msm_optimize} or a Chapman-Kolmogorov test \cite{CK_test} to help ensure self-consistency, these tests will not reveal whether key regions or pathways of the system have been missed. The authors are not aware of any analysis capable of doing this. Nevertheless, WE remains an unbiased sampling procedure, and this study shows that it leads to improved sampling compared to the initial unoptimized WE runs. 
    
    Will the WE optimization be worth the extra costs? The overhead costs consist of (a) obtaining training data, (b) building a haMSM model, and (c) determining the optimized bins from the haMSM.
    However, cost (a) is part of the standard workflow, since users are likely to run initial WE simulations using intuitive binning, which either will prove sufficient for the desired goals or not. Cost (b) entails approximately 2 days of cluster computing to process 30 $\mu$s of molecular time, as shown in Table \ref{table:computational_cost}.
    This cost will increase modestly with system size when using standard dimensionality reduction techniques.
    For some users, however, this cost is effectively free since they may wish to construct a haMSM for analysis purposes anyway.
    Last, cost (c) requires less than a second of computation using a pre-calculated hash table of cluster and bin indices, and this also will scale weakly with the system size. Compared to costs (b) and (c), the cost for WE simulation (optimized or not) is much larger, often days--weeks on a computing cluster.

    To help determine whether or not WE bin optimization is worthwhile, future studies could use the haMSM model to estimate the ``variance improvement constant'' of WE simulations as discussed in our prior work.\cite{WE_recent_mathematical_developments} In some systems, the maximum improvement to the variance may be modest, and this can be estimated by comparing the variance constants for optimized versus unoptimized WE bins. 
    Further, to help optimize cluster construction, the variance constants could be evaluated for different clustering methods and hyperparameters.


    Our work points to several possible improvements for the WE optimization method. First, given a certain amount of training data, there may be better ways to estimate the $h$ and $v$ functions used in the optimization than building the haMSM model directly. Since the initial data likely does not conform to the true steady state, recently proposed approaches for reweighting into steady state\cite{RiteWeight} could enable more accurate haMSM models from the same data. Also, one could iteratively improve $h$ and $v$ estimates as more data is gathered in a series of optimized runs.
    Next, as a different type of improvement, our computational pipeline could be made more efficient. In the present implementation using haMSMs, featurization and cluster assignment are computationally intensive, and they could be optimized by using GPU resources.
    Last, we note that explicit consideration of velocity coordinates may be required to adequately sample underdamped systems \cite{vanerp2012dynamical}.

\section{6. Conclusion}

    In this work, we applied a previously developed framework\cite{WE_recent_mathematical_developments} for reducing the run-to-run variance of weighted-ensemble (WE) estimates of mean first-passage times (MFPTs). The approach uses initial simulation data to estimate the local MFPT values in order to construct optimized WE bins that are kinetically similar. To test the effectiveness of the approach, we studied a synthetic MD model for Trp-cage folding and unfolding along with two atomistic models of NTL9 folding. Overall, we found that the high-friction systems with slow relaxation times were most amenable to improvement. Extremely encouraging results were obtained for the most challenging system studied, atomistic NTL9 in implicit solvent with a high, water-like friction constant. Whereas only 70\% of unoptimized WE runs for high-friction NTL9 yielded successful folding events, every optimized run yielded events and consistent MFPT estimates. 
    
\begin{acknowledgement}

This work was supported by NIH grant GM115805 to D. Zuckerman. Additionally, the research reported in this publication used computational infrastructure supported by the Office of Research Infrastructure Programs, Office of the Director, of the National Institutes of Health under Award Number S10OD034224.
D. Aristoff and G. Simpson gratefully acknowledge the support from the National Science Foundation via Award No. DMS 2111278.

\end{acknowledgement}

\begin{suppinfo}

The synMD package and the Trp-cage model can be found at \url{https://github.com/jdrusso/SynD}. A Python script to build an haMSM model and run optimized WE binning can be found at \url{https://github.com/ZuckermanLab/WE_variance_minimization}. An updated package to perform haMSM modeling can be found at \url{https://github.com/ZuckermanLab/msm_we}. 

\end{suppinfo}

\bibliography{manuscript}

\end{document}



\newpage
\section{1. Low Friction NTL9 Optimization Functions} 

\begin{figure}
    \centering
    \includegraphics[width=\linewidth]{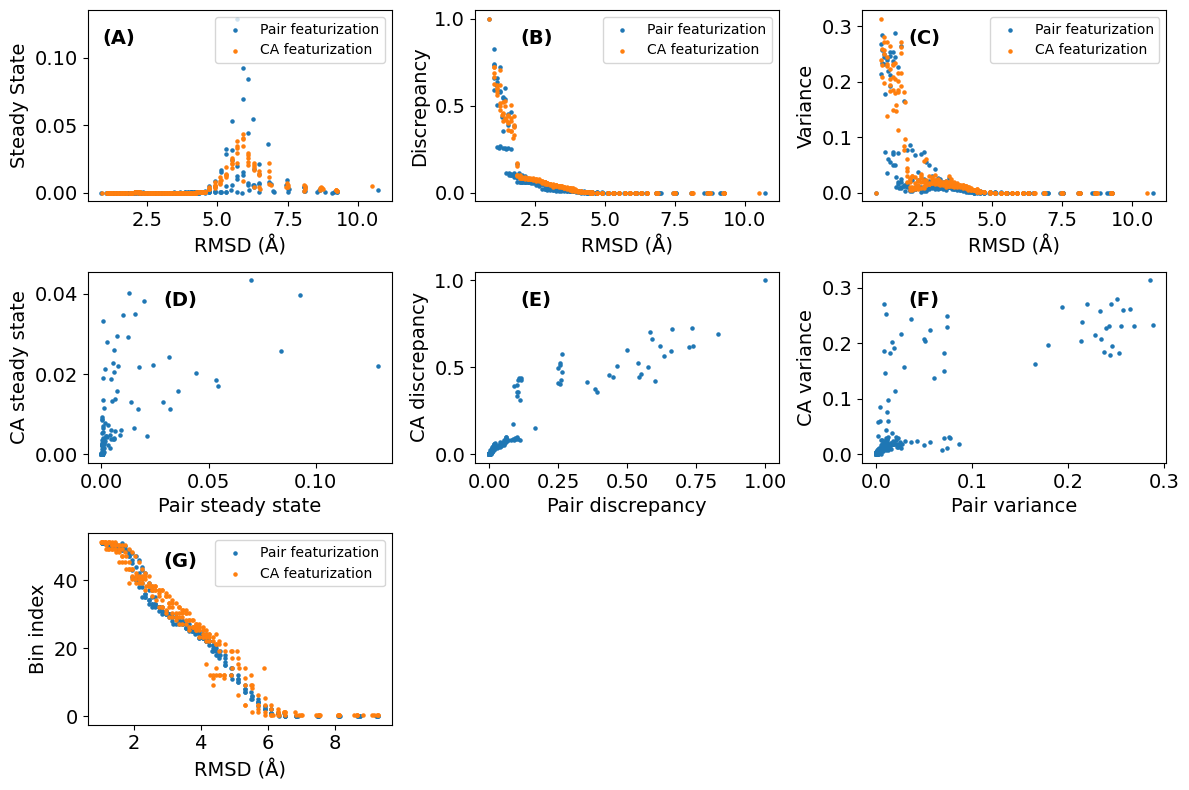}
    \caption{haMSM results for the low friction NTL9 haMSM model. Data for two models, using C$\alpha$ distance and pair distance featurization as discussed in the main text, are shown. (A) Steady state estimate, (B) discrepancy estimate, (C) variance estimate, (D) scatter plot of C$\alpha$ featurization steady state and pair featurization steady state, (E) scatter plot of the two discrepancy functions, (F) scatter plot of the two variance functions, and (G) optimized bin index as a function of RMSD for the two haMSM models.} 
    \label{fig:ntl9_low_haMSM}
\end{figure}

\hspace*{1cm}\\

\newpage
\section{2. High Friction NTL9 Optimization Functions}

\begin{figure}
    \centering
    \includegraphics[width=\linewidth]{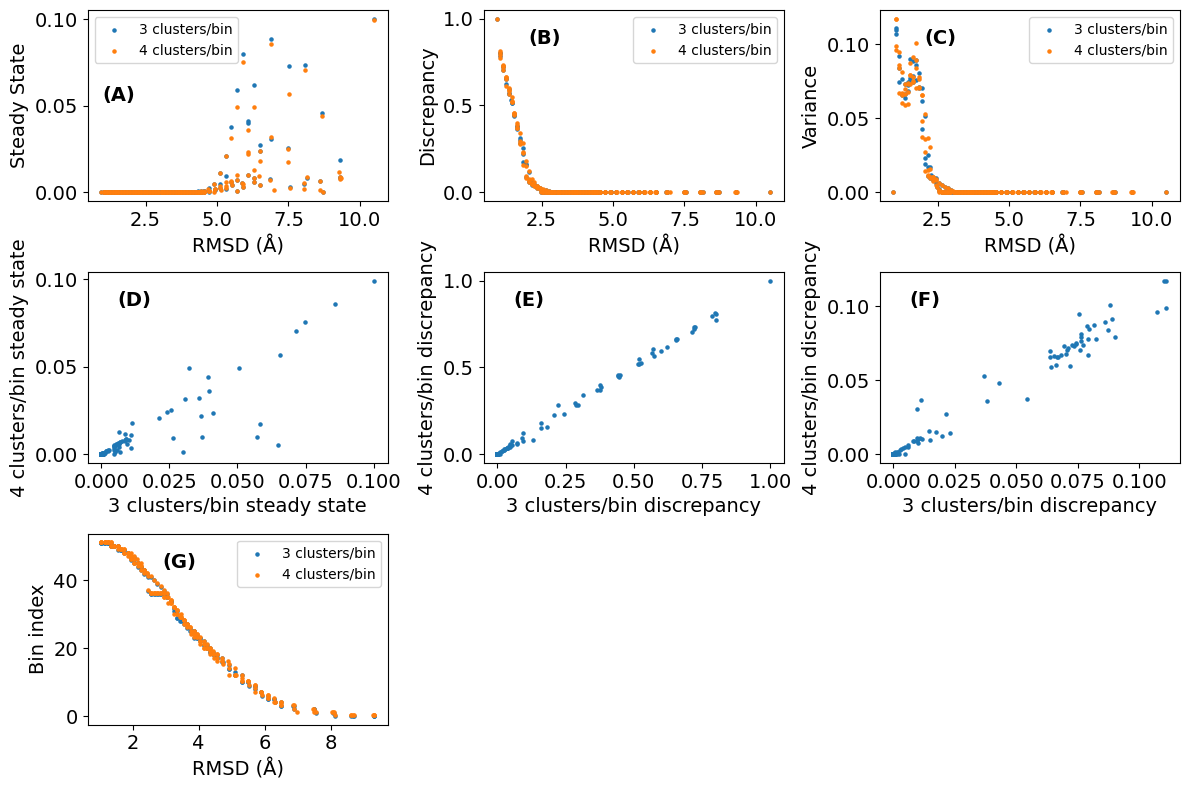}
    \caption{haMSM results for the high friction NTL9 haMSM model. Data for two models, using 3 clusters/bin and 4 clusters/bin as discussed in the main text, are shown. (A) Steady state estimate, (B) discrepancy estimate, (C) variance estimate, (D) scatter plot of 3 clusters/bin featurization steady state and 4 clusters/bin steady state, (E) scatter plot of the two discrepancy functions, (F) scatter plot of the two variance functions, and (G) optimized bin index as a function of RMSD for the two haMSM models.} 
    \label{fig:ntl9_high_haMSM}
\end{figure}